\documentclass[runningheads,citeauthoryear]{apinv_kurtedit}
\usepackage{epsfig,cite,graphics}
\usepackage{marvosym} %For astronomical symbols % % %
\usepackage[utf8]{inputenc}

\def\kms{km~s$^{-1}$} 

\begin{document}

\title{Optical flickering of the symbiotic star CH Cyg}
\titlerunning{Flickering of the symbiotic star CH Cyg}
\author{K. A. Stoyanov\inst{1}, J. Mart\'i\inst{2}, R. Zamanov\inst{1}, V. V. 
Dimitrov\inst{1}, A. Kurtenkov\inst{1}, E. S\'anchez-Ayaso\inst{2}, I. 
Bujalance-Fern\'andez\inst{2}, G. Y. Latev\inst{1}, G. Nikolov\inst{1}}
\authorrunning{Stoyanov, Mart\'i, Zamanov et al.}
\tocauthor{K. A. Stoyanov, J. Mart\'i, R. Zamanov, G. Y. Latev}
\institute{Institute of Astronomy and National Astronomical Observatory, 
Bulgarian Academy of Sciences, 72 Tsarigradsko Chaussee Blvd., 1784, Sofia, 
Bulgaria
    \and Departamento de F\'isica (EPSJ), Universidad de Ja\'en, Campus Las 
Lagunillas, A3-420, 23071, Ja\'en, Spain \newline
    \email{kstoyanov@astro.bas.bg jmarti@ujaen.es rkz@astro.bas.bg}}
\papertype{Submitted on 29 Aug 2017; Accepted on xx.xx.xxxx}	
% Papertype can be "Research report", "Review", "Invited lecture", "Conference talk", 
% "Conference poster", "Lecture at scientific seminar", "Summary of 
%dissertation",  etc.
\maketitle

\begin{abstract}
Here we present quasi-simultaneous observations of the flickering of the 
symbiotic binary star CH Cyg in U, B and V bands. We calculate the flickering 
source parameters and discuss the possible reason for the flickering cessation 
in the period 2010--2013.
\end{abstract}
\keywords{stars: binaries: symbiotic -- accretion, accretion discs  -- stars: 
individual: CH~Cyg}

\section{Introduction}
Symbiotic stars are interacting binaries. They consist of an evolved red giant 
or Mira-type variable, and a hot component --- white dwarf, subdwarf, neutron 
star or main-sequence star. The hot component accretes material from the stellar 
wind of the donor (Sokoloski 2003). The wind is ionized by the hot component, 
causing the rise of a nebula. Orbital periods of the symbiotic stars range from 
years to decades.

CH~Cyg is an eclipsing symbiotic star composed of a M6-7~III star and an 
accreting white dwarf, so the system belongs to the S-type symbiotics. The 
binary separation is $8.7^{+1.1}_{-0.7}$~AU (Miko{\l}ajewska et al., 2010). The 
masses of the components are $M_{\rm rg}=2^{+1}_{-0.5}\, \rm M_{\sun}$ and 
$M_{\rm wd}=0.70^{+0.22}_{-0.09}\, \rm M_{\sun}$ (Miko{\l}ajewska et al., 2010). 
Based on {\it Hipparcos} satellite measurements, the distance to CH~Cyg is 
estimated to $d=244^{+49}_{-35}\, \rm pc$ (van Leeuven 2007). CH~Cyg ejects 
collimated bipolar outflows with velocity of $\sim$700~\kms, detectable in the 
radio (Taylor, Seaquist \& Mattei 1986; Crocker et al., 2001). The system is 
detectable also in X-rays (Galloway \& Sokoloski 2004). The orbital period of 
CH~Cyg is $\sim$15.6~yr (Hinkle, Fekel \& Joyce 2009). The light curve of CH~Cyg 
is very complex. The detected variability varies from dozens of years caused by 
the orbital motion and dust obscuration events (Bogdanov \& Taranova 2001), 
through periodicity with periods of several hundred days caused by pulsations of 
the giant (Mikolajewski, Mikolajewska \& Khudyakova 1992), to flickering 
activity with time-scales of a few minutes (Dobrzycka, Kenyon \& Milone 1996).

Flickering is broad-band stochastic light variations on time-scales of a few 
minutes with amplitude from a few$\times$0.01~mag to more than one magnitude. 
Flickering activity is detected in only 10 symbiotic stars --- RS~Oph, T~CrB, 
MWC~560, V2116~Oph, CH~Cyg, RT~Cru, $o$~Cet, V407~Cyg, V648~Car and EF~Aql 
(Dobrzycka, Kenyon \& Milone 1996; Sokoloski, Bildsten \& Ho  2001; Gromadzki et 
al. 2006; Angeloni et al. 2012; 
Zamanov et al., 2017).

The flickering of CH~Cyg was first detected by Wallerstein (1968) and Cester 
(1968) and it was studied in detail later (Skopal 1988; Mikolajewski et al. 
1990; Mikolajewski et al. 1992; Panov \& Ivanova 1992; Kuczawska, Mikolajewski 
\& Kirejczyk 1992; Hric et al. 1993; Dobrzycka, Kenyon \& Milone 1996; Sokoloski 
\& Kenyon 2003). CH~Cyg is an eclipsing system (Mikolajewski, Mikolajewska \& 
Tomov 1987), and the flickering activity disappears during the eclipses 
(Sokoloski \& Kenyon 2003). In 2010, the flickering from  CH~Cyg became 
non-detectable (Sokoloski et al., 2010) until it renewed its activity in 2014 
(Stoyanov et al., 2014). CH~Cyg probably enters a new active stage in 2017 
(Iijima 2017).

Here we present photometric observations of the flickering of CH Cyg and 
calculations of the flickering source parameters.

\section{Observations}

The observations are performed with the following three telescopes equipped with 
CCD cameras:
\begin{itemize}
  \item the 60 cm Cassegrain telescope of Rozhen NAO
  
  \item the 50/70 cm Schmidt telescope of Rozhen NAO 
  
  \item the automated 41 cm telescope of the University of Ja\'en, Spain 
(Mart{\'{\i}}, Luque-Escamilla, \& Garc\'{\i}a-Hern\'andez 2017)
% , 0.42 arcsec px$^{-1}$). 
\end{itemize}

In Fig.~\ref{fig.obs} are plotted the light curves from a few nights. The 
observations consisted of repeated exposures in U, B and V bands, or in B and V 
bands. On 20110609 the total duration of the run is 86 minutes; 20141001 --- 90 
minutes; 20170724 --- 66 minutes; 20170809 --- 147 minutes; 20170811 --- 284 
minutes.

The data reduction was done using IRAF (Tody 1993) following standard procedures 
for aperture photometry. A few comparison stars from the list of Henden \& 
Munari (2006) have been used, bearing in mind that SAO~31628 is an eclipsing 
binary (Sokoloski \& Stone 2000).

The journal of observations is given in Table~\ref{table1}. In the table are 
given the telescope, band, number of exposures, exposure time, average 
magnitude, minimum and maximum magnitude during the run, and typical 
observational error.

\section{Flickering source parameters}

\begin{figure*}    
   \vspace{16.0cm}   
   \includegraphics{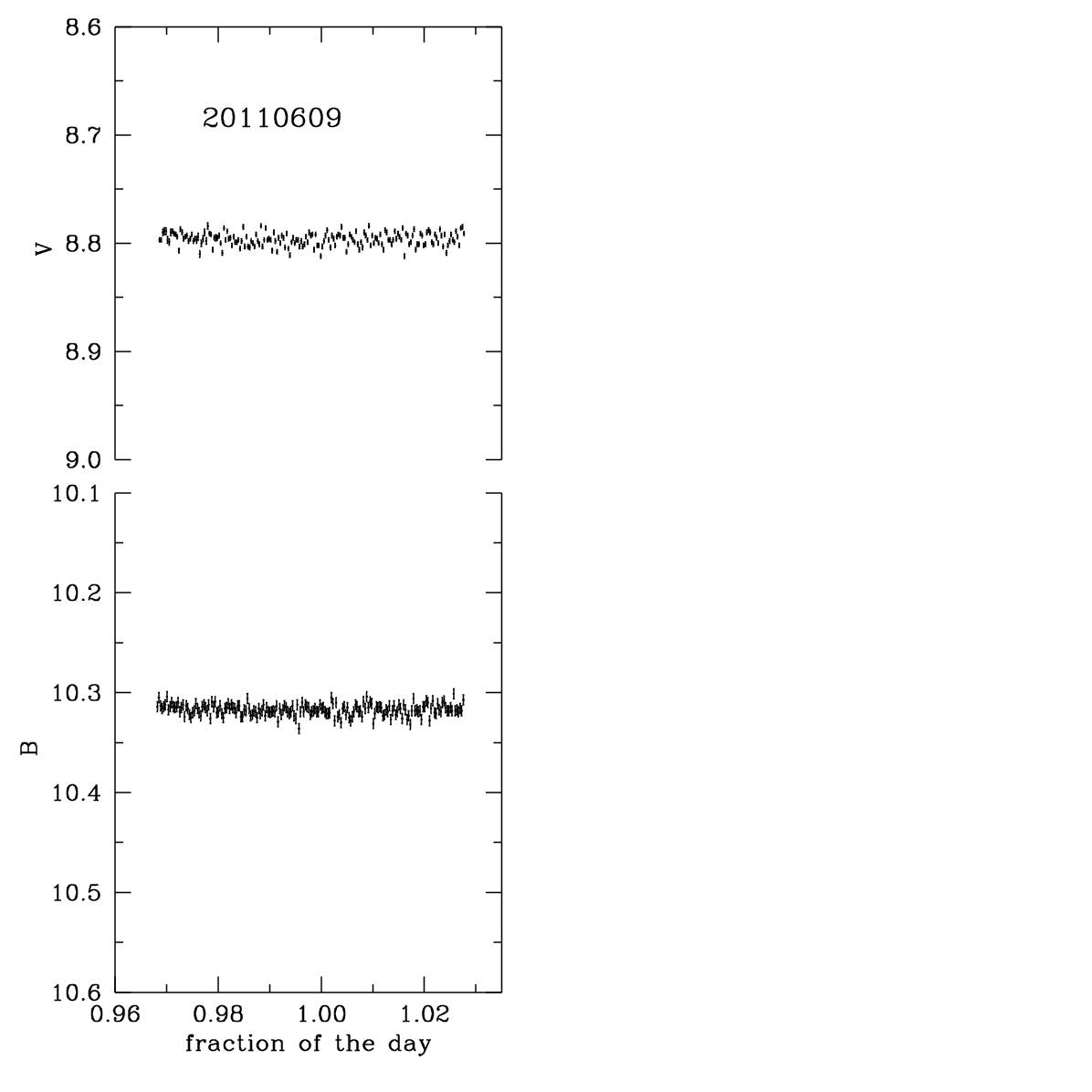}    
   \includegraphics{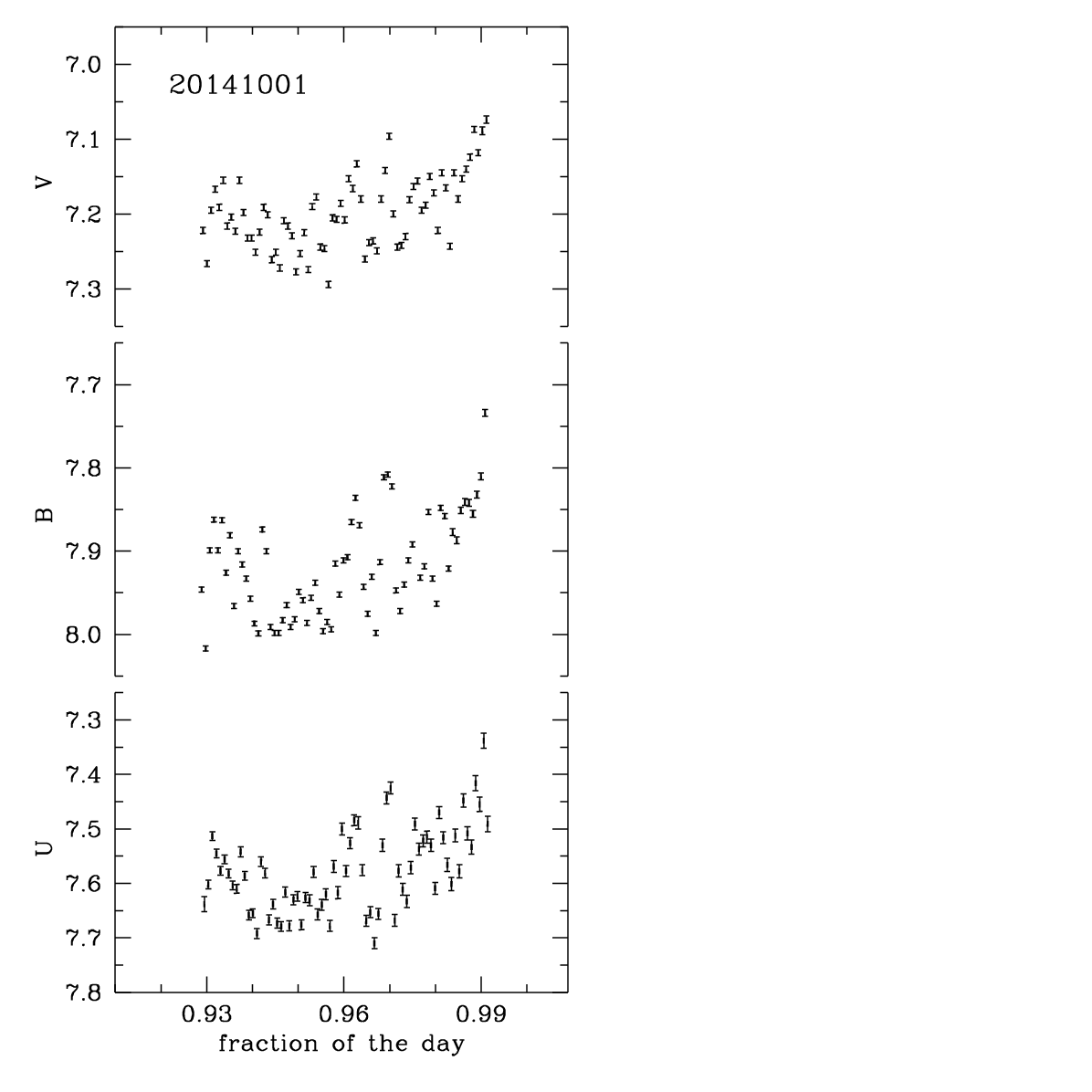}  
   \includegraphics{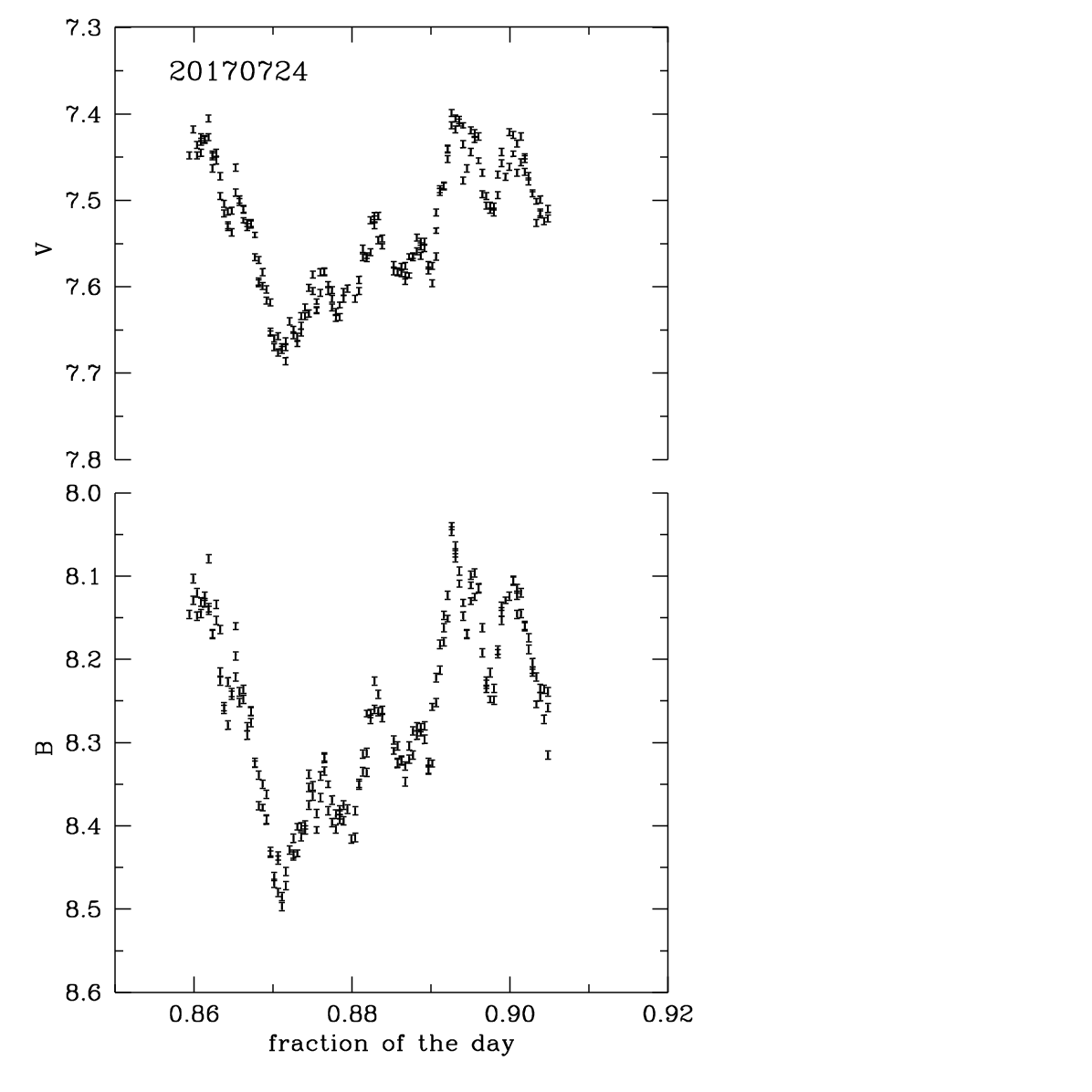}      
   \includegraphics{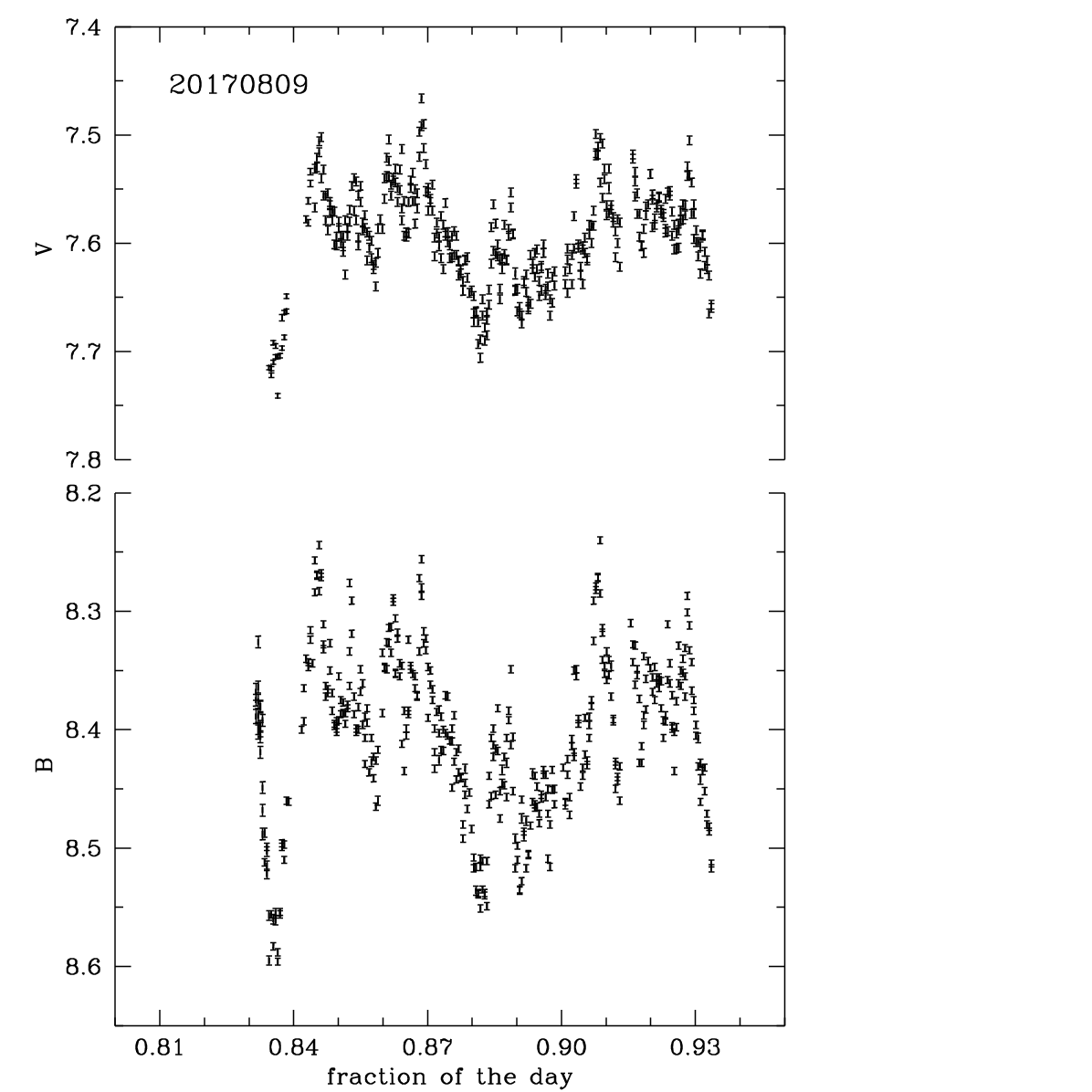}      
   \includegraphics{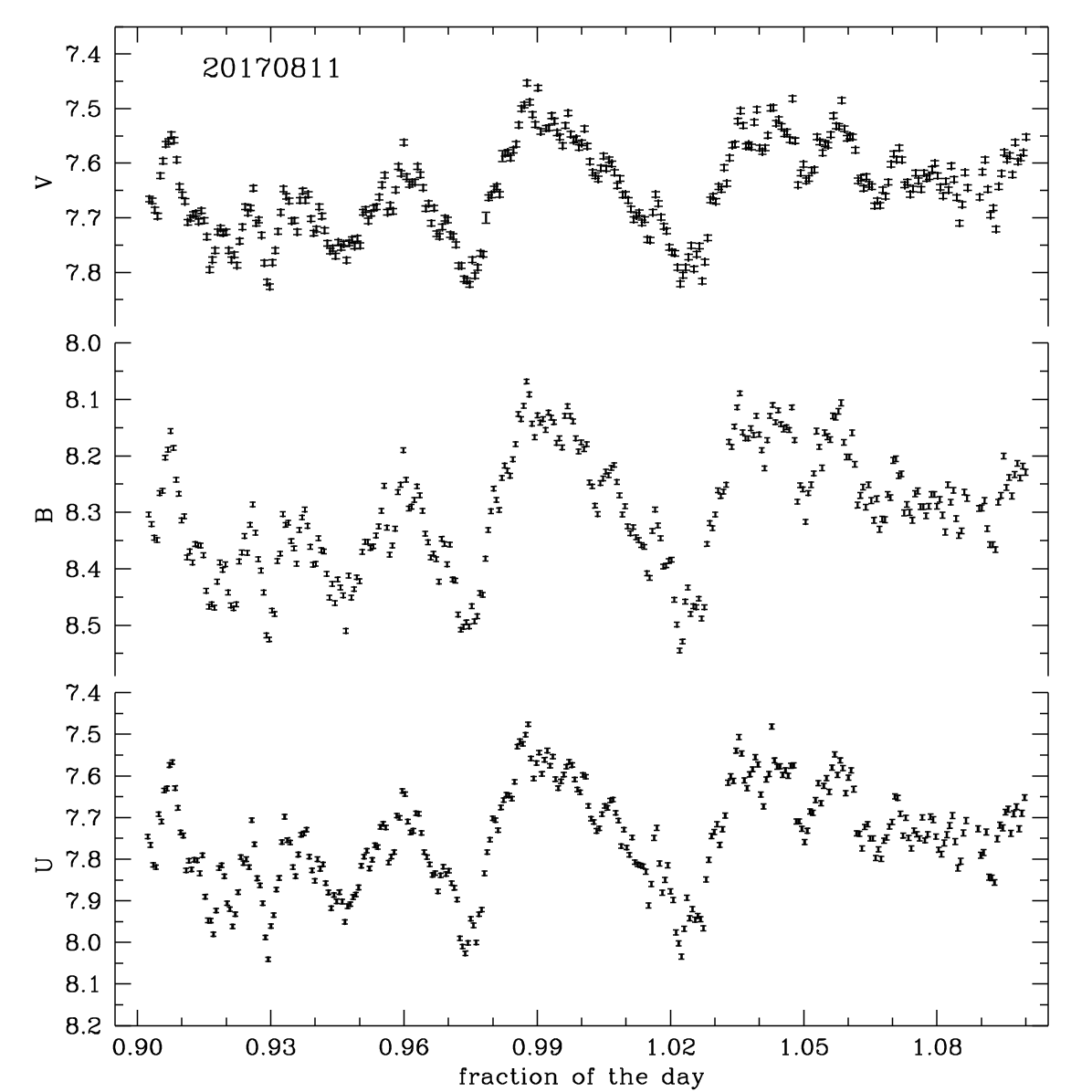}      
   \caption[]{Flickering of CH~Cyg in U, B and V bands.}
   \label{fig.obs}           
\end{figure*}

In our observations obtained during the period 2010--2013 the flickering of 
CH~Cyg was not detectable (see Table~\ref{flick} and Fig.~\ref{fig.obs}). It 
re-appeared in August 2014. After that CH~Cyg exhibited variability on a time 
scale of 1--30 minutes with amplitude $0.2-0.3$~mag in $V$. The amplitude 
increases in B and U bands.

Bruch (1992) proposed that the light curve of CVs can be separated into two 
parts --- constant light, and variable (flickering) source. We assume that all 
the variability in each night is due to flickering. In these suppositions the 
flickering light source is considered 100\% modulated. Following these 
assumptions, we calculate the flux of the flickering light source as $F_{\rm 
fl}=F_{\rm av}-F_{\rm min}$, where $F_{\rm av}$ is the average flux during the 
run and $F_{\rm min}$ is the minimum flux during the run (corrected for the 
typical error of the observations). $F_{\rm fl}$ has been calculated for each 
band, using the values given in Table~1 and the Bessel (1979) calibration for 
the fluxes of a zero magnitude star.

A modification of this method is given in Nelson et al. (2011), which proposed 
to use the $F_{\rm fl}=F_{\rm max}-F_{\rm min}$, where $F_{\rm max}$ is the 
maximum flux during the run. Adopting these, we find that the flickering light 
source contributes about 4\% in $V$, 6\% in $B$,  and 8\% in $U$ (October 2014).

The calculated colours of the flickering light source are given in 
Table~\ref{table2}, where $T_{(B-V)_1}$ is calculated using $F_{av}$ and 
$T_{(B-V)_2}$ is calculated using $F_{max}$.

Assuming that the flickering source radiates as a black body, for the colours of 
the black body we use the calibration given in Strai\v zis (1977). The use of 
other formulae (e.g. Ballesteros 2012) could introduce a difference of about 
$\pm 500$~K.

An independent estimate of the black body temperatures from the B-V colour using 
an analytic approximation (Ballesteros 2012) provides similar results with a 
difference not exceeding 1000 K.

The radius is calculated for B-V colour and B band flux, assuming effective 
wavelength of B band $\lambda = 4400$~\AA, the temperature calculated from the 
B-V colour, assuming black body, spherical form of the flickering source, and a 
distance $d=244 pc$.

In a comparison of the temperatures, calculated using the method of Bruch (1992) 
and Nelson et al. (2011), we find that the temperatures are in agreement.

As expected, the method of Nelson et al. (2011) gives higher values for the size 
of the flickering source, because it uses greater flux.

In Table~\ref{table2} are summarized the calculated flickering source 
parameters: 
    (1) date of observations; 
    (2) U-B colour of the flickering source calculated following Bruch (1992); 
    (3) temperature corresponding to the U-B colour;
    (4) B-V colour of the flickering source calculated following Bruch (1992); 
    (5) temperature corresponding to B-V colour;
    (6) radius of the flickering source calculated following Bruch (1992);
    (7) U-B colour of the flickering source calculated following Nelson et al. 
(2011); 
    (8) temperature corresponding to the U-B colour;
    (9) B-V colour of the flickering source calculated following Nelson et al. 
(2011); 
    (10) temperature corresponding to B-V colour;
    (11) radius of the flickering source calculated following Nelson et al. 
(2011).

%------------------------------------------------------------------------
%--- 20170809.J.eps--------------------------------------------------------

%--------------------------------------------------------------------------
\begin{table}
   \begin{center}
   \caption{Journal of observations.}
\begin{tabular}{l cccc  ccccc cc}
date-obs & telescope & band & exposures & average & min & max & merr & \\
\\
20110609 & 60cm Roz &    B & 191 $\times$ 10s &  10.317 &  10.301 & 10.336 & 
0.005 & \\
20110609 & 60cm Roz &    V & 190 $\times$ 5s  &   8.796 &   8.784 &  8.812 & 
0.002 & \\
\\
20141001 & 60cm Roz &    U &  69 $\times$ 15s &   7.578 &   7.344 &  7.714 & 
0.009 & \\
20141001 & 60cm Roz &    B &  71 $\times$ 10s &   7.916 &   7.735 &  8.017 & 
0.003 & \\
20141001 & 60cm Roz &    V &  71 $\times$ 4s  &   7.197 &   7.074 &  7.294 & 
0.004 & \\
\\
20170724 & 41cm Jaen &   B &  200 $\times$ 3s &   8.264 &   8.040 &  8.497 & 
0.009 & \\
20170724 & 41cm Jaen &   V &  196 $\times$ 2s &   7.533 &   7.399 &  7.686 & 
0.005 & \\
\\
20170809 & 41cm Jaen &   B &  417 $\times$ 6s &   8.410 &   8.248 &  8.600 & 
0.003 & \\ 
20170809 & 41cm Jaen &   V &  389 $\times$ 2s &   7.666 &   7.547 &  7.827 & 
0.004 & \\ 
\\
20170811 &50/70cm Roz&   U & 318 $\times$ 20s &   7.748 &   7.476 &  8.041 & 
0.007 & \\
20170811 &50/70cm Roz&   B & 318 $\times$ 4s  &   8.298 &   8.068 &  8.545 & 
0.005 & \\ 
20170811 &50/70cm Roz&   V & 318 $\times$ 2s  &   7.649 &   7.453 &  7.826 & 
0.005 & \\
\\
\end{tabular}
\label{table1}
\end{center}
\end{table}
%-----------------------------------------------------------

%-----------------------------------------------------------
\begin{table}
\caption{Flickering source parameters.}
\resizebox{\textwidth}{!}{
\begin{tabular}{c | cc | ccc | cc | ccccc }
date-obs & $U-B$ & $T_{(U-B)_1}$ & $B-V$ & $T_{(B-V)_1}$ & $R/R_\odot$ & $U-B$ &
$T_{(U-B)_2}$ & $B-V$ & $T_{(B-V)_2}$ & $R/R_\odot$ & \\
& & & & & & & & & & & \\
1 & 2 & 3 & 4 & 5 & 6 & 7 & 8 & 9 & 10 & 11 & \\
& & & & & & & & & & & \\
20141001 & -0.6439 & 9292 & 0.6782 & 5554 & 1.94 & -0.6441 & 9284 & 0.4212 & 
6992 &
2.07 & \\
20170724 & --- & --- & 0.3126 & 7895 & 1.20 & --- & --- & 0.2161 & 9049 & 1.42 & 
\\
20170809 & --- & --- & 0.5430 & 6231 & 1.62 & --- & --- & 0.5322 & 6299 & 2.24 & 
\\
20170811 & -0.7134 & 9834 & 0.3211 & 7824 & 1.23 & -0.7353 & 10723 & 0.3966 & 
7195 &
2.10 & \\
\end{tabular}
}
\label{table2}
\end{table}
%-----------------------------------------------------------

%-----------------------------------------------------------
\begin{table}
\begin{center}
\caption{List of observations of CH~Cyg. The last column indicates if the 
flickering
is present or not.}
\begin{tabular}{c | c c | c | c | c | c | c | c c c c}
date-obs & bands & flickering & date-obs & bands & flickering \\
& & & & & \\
20100430 & BV & no & 20110918 & B & no \\ 
20100501 & UB & no & 20111006 & B & no \\
20100502 & B & no & 20111118 & B & no \\
20100506 & B & no & 20120618 & B & no \\
20100507 & UB & no & 20120820 & B & no \\
20100509 & BV & no & 20130514 & BV & no \\
20100816 & UB & no & 20130703 & B & no \\
20100817 & BV & no & 20130803 & B & no \\
20100818 & UB & no & 20140814 & B & yes \\
20100909 & BV & no & 20141001 & UB & yes \\
20101029 & UB & no & 20170724 & BV & yes \\
20101030 & BV & no & 20170809 & BV & yes \\
20110529 & B & no & 20170811 & UBV & yes \\
20110609 & BV & no & & &\\
\end{tabular}
\label{flick}
\end{center}
\end{table}
%-----------------------------------------------------------

\section{Discussion}

\subsection*{Parameters of the flickering source}

The temperature of the flickering source in similar to the temperature of the 
bright spot in cataclysmic variable stars (e. g. Marsh 1988;
Wood et al., 1989; Zhang \& Robinson 1987) and it is significantly lower than 
the temperature of the boundary layer. This is a hint that the flickering 
probably originates from the bright spot, but its nature is still unknown. For 
comparison, the temperatures and radii of the flickering source in other 
symbiotic stars are comparable with those that we estimate for CH~Cyg. For 
RS~Oph, Zamanov et al. (2010) give $T_{fl}=9500 \pm 500$~K and $R_{fl} = 3.5 \pm 
0.5 R_\odot$. For MWC~560, Zamanov et al. (2011) give $T_{fl}=13550 \pm 500$~K 
and $R_{fl} = 1.68 \pm 0.16 R_\odot$.

\subsection*{Why was the flickering missing for 4 years?}

Flickering is expected to arise in the vicinity of the accretion disc around the 
white dwarf companion in CH~Cyg. 
Its absence during a nearly four year time interval is most naturally 
interpreted as a major disruption of the inner disc structure. 
This is probably due to reduced supply of mass flow from the M-type giant across 
the L$_1$ Lagrangian point of the system. 
This situation, lasting in the period 2010--2013 according to Table~\ref{flick}, 
renders now difficult other alternative interpretations 
for the lack of flickering based on an eclipse configuration (Stoyanov et al. 
2014). The required eclipse duration, at 
least four years, appears to be too long unless a highly eccentric system is 
invoked. Interestingly, in the CH~Cyg case there 
is observational evidence based on infrared observations for episodic creation 
and dissipation of a dust envelope around it 
(Taranova \& Shenavrin 2004). Such a behavior has been observed in the past. 
The creation and dissipation of a dust envelope is connected with a change of the mass-loss of the giant, which 
underwent a significant reduction during the 
2010--2013 time interval. The time scale reported by 
Taranova \& Shenavrin (2004) for previous dust envelope creation and dissipation 
 events is $\sim$ 10 yr. The cessation of the flickering is consistent with 
the giant undergoing a reduced mass-loss episode with a shorter, few year 
duration. Unfortunately, no contemporaneous infrared 
observations are available to us to confirm this hypothesis.

%CH CYGNI II: OPTICAL FLICKERING FROM AN UNSTABLE DISK    J. L. Sokoloski and S. J. Kenyon

%Sokoloski, J. L.; Kenyon, S. J. CH Cygni. I. Observational Evidence for a Disk-Jet Connection

%Flickering source parameter

%Why the flickering was missing for 4 years? 

%1996ASSL..208..341M  Mikolajewski, M.; Tomov, T.; Dapergolas, A.; Bellas-Velidis, Y.	
%Possible periodic components in the flickering of CH CYG and MWC 560

%The ingress duration was less than one day. The radius of the red giant is
%estimated to be $288 \pm 15$ R from the duration of the eclipse.
% the duration of the eclipse, $94 \pm 5$ d  Iijima, 1998, MNRAS.  

\section*{Conclusions}

We performed quasi-simultaneous multicolour observations of the flickering of 
the symbiotic star CH Cyg in Johnson U, B and V bands. We calculated the 
flickering source parameters --- the temperature is in the range $5000 < T < 
11~000$~K and the radius is in the range 1.42 $<R/R_\odot<$2.24. We briefly 
discuss the disappearance and re-appearance of the flickering.

\vskip 0.5cm 

\noindent
{\bf Acknowledgments:} We thank the referee, N. Tomov, for his constructive 
comments and suggestions. This work was partially supported by the Bulgarian 
National Science Fund of the Ministry of Education and Science under grant 
DN~08-1/2016, and AYA2016-76012-C3-3-P from the Spanish Ministerio de Econom\'ia 
y Competitividad (MINECO), and by the Consejer\'ia de Econom\'ia, Innovaci\'on, 
Ciencia y Empleo of Junta de Andaluc\'ia under research group FQM-322, as well 
as FEDER funds. 

%\newpage


\begin{thebibliography}{}
\bibitem{2012ApJ...756L..21A} Angeloni, R., Di Mille, F., 
Ferreira Lopes, C.~E., \& Masetti, N.\ 2012, \apjl, 756, L21 

\bibitem{2012EL.....9734008B} Ballesteros, F.~J.\ 2012, EPL 
(Europhysics Letters), 97, 34008 

\bibitem{1979PASP...91..589B} Bessell, M.~S.\ 1979, \pasp, 91, 
589 

\bibitem{2001ARep...45..797B} Bogdanov, M.~B., \& 
Taranova, O.~G.\ 2001, Astronomy Reports, 45, 797 

\bibitem{1992A&A...266..237B} Bruch, A.\ 1992, \aap, 266, 237 

\bibitem{1968IBVS..291....1C} Cester, B.\ 1968, Information 
Bulletin on Variable Stars, 291, 1 

\bibitem{2001MNRAS.326..781C} Crocker, M.~M., Davis, 
R.~J., Eyres, S.~P.~S., et al.\ 2001, \mnras, 326, 781 

\bibitem{1996AJ....111..414D} Dobrzycka, D., Kenyon, 
S.~J., \& Milone, A.~A.~E.\ 1996, \aj, 111, 414 

\bibitem{2004ApJ...613L..61G} Galloway, D.~K., \& 
Sokoloski, J.~L.\ 2004, \apjl, 613, L61 

\bibitem{2006AcA....56...97G} Gromadzki, M., 
Mikolajewski, M., Tomov, T., et al.\ 2006, \actaa, 56, 97 

\bibitem{2006A&A...458..339H} Henden, A., \& Munari, U.\ 
2006, \aap, 458, 339 

\bibitem{2009AAS...21343410H} Hinkle, K.~H., Fekel, F.~C., 
\& Joyce, R.~R.\ 2009, Bulletin of the American Astronomical Society, 41, 
434.10 

\bibitem{1993CoSka..23...73H} Hric, V., Skopal, A., Urban, 
Z., et al.\ 1993, Contributions of the Astronomical Observatory Skalnate Pleso, 
23, 73 

\bibitem{2017ATel.10142...1I} Iijima, T.\ 2017, The Astronomer's 
Telegram, 1014,  

\bibitem{1992IBVS.3806....1K} Kuczawska, E., 
Mikolajewski, M., \& Kirejczyk, K.\ 1992, Information Bulletin on Variable 
Stars, 3806, 1 

\bibitem{1988MNRAS.231.1117M} Marsh, T.~R.\ 1988, \mnras, 231, 
1117 

\bibitem{2017BlgAJ..26...91M} Mart{\'{\i}}, J., 
Luque-Escamilla, P.~L., \& Garc{\'{\i}}a-Hern{\'a}ndez, M.~T.\ 2017, Bulgarian 
Astronomical Journal, 26, 91 

\bibitem{2010MNRAS.403L..21M} Miko{\l}ajewska, J., 
Balega, Y., Hofmann, K.-H., \& Weigelt, G.\ 2010, \mnras, 403, L21 

\bibitem{1992A&A...254..127M} Mikolajewski, M., 
Mikolajewska, J., \& Khudyakova, T.~N.\ 1992, \aap, 254, 127 

\bibitem{1987Ap&SS.131..733M} Mikolajewski, M., 
Mikolajewska, J., \& Tomov, T.\ 1987, \apss, 131, 733 

\bibitem{1990AcA....40..129M} Mikolajewski, M., 
Mikolajewska, J., Tomov, T., Kulesza, B., \& Szczerba, R.\ 1990, \actaa, 40, 
129 

\bibitem{1992IBVS.3742....1M} Mikolajewski, M., 
Tomov, T., Kuczawska, E., et al.\ 1992, Information Bulletin on Variable Stars, 
3742, 1 

\bibitem{2011ApJ...737....7N} Nelson, T., Mukai, K., Orio, 
M., Luna, G.~J.~M., \& Sokoloski, J.~L.\ 2011, \apj, 737, 7 

\bibitem{1992IBVS.3817....1P} Panov, K.~P., \& Ivanova, 
M.~S.\ 1992, Information Bulletin on Variable Stars, 3817, 1 

\bibitem{1988ASSL..145..223S} Skopal, A.\ 1988, IAU Colloq.~103: 
The Symbiotic Phenomenon, 145, 223 

\bibitem{2003JAVSO..31...89S} Sokoloski, J.~L.\ 2003, Journal 
of the American Association of Variable Star Observers (JAAVSO), 31, 89 

\bibitem{2001MNRAS.326..553S} Sokoloski, J.~L., 
Bildsten, L., \& Ho, W.~C.~G.\ 2001, \mnras, 326, 553 

\bibitem{2003ApJ...584.1027S} Sokoloski, J.~L., \& 
Kenyon, S.~J.\ 2003, \apj, 584, 1027 

\bibitem{2000IBVS.4983....1S} Sokoloski, J.~L., \& 
Stone, R.~P.~S.\ 2000, Information Bulletin on Variable Stars, 4983, 1 

\bibitem{2010ATel.2707....1S} Sokoloski, J.~L., Zamanov, 
R., Stoyanov, K., Bryson, S., \& Still, M.\ 2010, The Astronomer's Telegram, 
2707,  

\bibitem{2014ATel.6560....1S} Stoyanov, K., Latev, G., 
Nikolov, G., Zamanov, R., \& Sokoloski, J.~L.\ 2014, The Astronomer's Telegram, 
6560,  

\bibitem{} Strai\v zis V., 1977, Multicolour Stelllar photometry, Mokslas 
Publishers, Vilnius, Lithuania

\bibitem{1986Natur.319...38T} Taylor, A.~R., Seaquist, 
E.~R., \& Mattei, J.~A.\ 1986, \nat, 319, 38 

\bibitem{2004ARep...48..813T} Taranova, O.~G., 
\& Shenavrin, V.~I.\ 2004, Astronomy Reports, 48, 813

\bibitem{1993nnnnn.xxx...xxx} Tody, D. 1993, "IRAF in the Nineties" 
in Astronomical Data Analysis Software
         and Systems II, A.S.P. Conference Ser., Vol 52, eds. R.J. Hanisch,
         R.J.V. Brissenden, \& J. Barnes, 173.

\bibitem{hippar07} van Leeuven, F.,  2007, Hipparcos, the New Reduction of 
the Raw Data (Heidelberg: Springer)

\bibitem{1968Obs....88..111W} Wallerstein, G.\ 1968, The 
Observatory, 88, 111 

\bibitem{1989ApJ...341..974W} Wood, J.~H., Horne, K., 
Berriman, G., \& Wade, R.~A.\ 1989, \apj, 341, 974 

\bibitem{2010MNRAS.404..381Z} Zamanov, R.~K., Boeva, S., 
Bachev, R., et al.\ 2010, \mnras, 404, 381 

\bibitem{2011IBVS.5995....1Z} Zamanov, R., Boeva, S., 
Latev, G., et al.\ 2011, Information Bulletin on Variable Stars, 5995, 1 

\bibitem{2017arXiv170208243Z} Zamanov, R.~K., Boeva, S., 
Nikolov, Y.~M., et al.\ 2017, Astronomische Nachrichten, 338, 680 

\bibitem{1987ApJ...321..813Z} Zhang, E.-H., \& 
Robinson, E.~L.\ 1987, \apj, 321, 813 

\end{thebibliography}
\end{document}